\documentclass[doublecol]{epl2} 

\usepackage{graphicx}
\usepackage[latin1]{inputenc}
\usepackage{graphicx,epstopdf}
\usepackage{dcolumn}
\usepackage{amsmath}
\usepackage[hypertex]{hyperref}
\usepackage{amsfonts}
\usepackage{amsthm}
\usepackage{amssymb}
\usepackage{mathrsfs}
\PassOptionsToPackage{caption=false}{subfig}
\usepackage{subfig}
\usepackage{color}
\usepackage[normalem]{ulem}

\title{Overcoming ambiguities in classical and quantum correlation measures}

\author{F. M. Paula\footnote{fagner@if.uff.br}\inst{1,2} \and A. Saguia\footnote{amen@if.uff.br}\inst{2} \and Thiago R. de Oliveira\footnote{tro@if.uff.br}\inst{2} \and M. S. Sarandy\footnote{msarandy@if.uff.br}\inst{2}}
\shortauthor{F. M. Paula \etal}

\institute{
  \inst{1} Universidade Tecnol\'ogica Federal do Paran\'a - Rua Cristo rei 19, Vila Becker, 85902-490, Toledo, PR, Brazil \\
  \inst{2} Instituto de F\'isica, Universidade Federal Fluminense - Av. Gal. Milton Tavares de Souza s/n, Gragoat\'a, 24210-346, Niter\'oi, RJ, Brazil}


\abstract{We identify ambiguities in the available frameworks for defining quantum, classical, and total correlations 
as measured by discordlike quantifiers. More specifically, we  determine  situations for which either classical 
or quantum correlations are not uniquely defined due to degeneracies arising from the optimization procedure over 
the state space. In order to remove such degeneracies, we introduce a general approach where correlations are 
independently defined, escaping therefore from a degenerate subspace. As an illustration, we analyze the trace-norm 
geometric quantum discord for two-qubit Bell-diagonal states.}

\begin{document}

\maketitle

\section{Introduction}

Quantum correlations are widely recognized as a resource for quantum information tasks~\cite{Modi}. 
In this scenario, entanglement plays a special role for applications in quantum computation and 
quantum communication~\cite{Nielsen-Chuang}. On the other hand, it is now known that, even in the 
absence of entanglement, it is possible to achieve some quantum advantage, such as in protocols 
for work extraction via Maxwell's demons~\cite{Zurek:03}, metrology~\cite{Modi:11,Girolami:13}, 
entanglement distribution~\cite{Chuan:12,Streltsov:12, Fedrizzi:13,Vollmer:13,Peuntinger:13}, quantum state 
merging~\cite{Madhok:11}, among others. The source for the quantum power of such tasks can be 
attributed to more general quantum correlations, as measured by quantum discord~\cite{Ollivier}. 
Such correlations can be suitably applied to make quantum systems supersede their classical counterparts. 

Quantum information science has then motivated the development of a general theory of quantum, classical, 
and total correlations in physical systems. In this context, quantum discord has been 
originally introduced by Ollivier and Zurek~\cite{Ollivier} as an entropic measure of quantum correlation 
in a bipartite system, which arises as a difference between the total correlation (as measured by the mutual 
information) before and after a local measurement is performed over one of the subsystems. In addition, a 
number of discordlike measures have also appeared through a geometric formulation, which are based, e.g.,  
on the relative entropy~\cite{Modi:10,Modi2:10}, Hilbert-Schmidt norm~\cite{Dakic:10,Bellomo:12}, trace 
norm~\cite{Paula,Nakano:12}, and Bures norm~\cite{Spehner:13,Bromley}. All of these distinct versions can 
be generally described by a unified framework in terms of a distance function. Here, the term 
distance will be generically used as a measure of distinguishability between two density operators (not 
necessarily a proper distance). Then, a correlation measure (either classical, quantum, or total) can be obtained 
by computing the distance function associated with a state $\rho$ and a related classical or product state.

By focusing on the quantumness of $\rho$, we typically optimize the distance function via a pre-selected
strategy over classical states, which leads to a unique value for the quantum correlation. We then compute the 
distance function between the optimal classical state and a product state (or a measured product state) to obtain 
the classical correlation. However, as we will explicitly show in this work, there may be more than a single optimal 
classical state. This set of optimal classical states, which are degenerate in the sense that they provide the same 
value for the quantum correlation, can lead to a nonunique (multivalued) classical correlation. Therefore, there is an  
ambiguity in the definition of the classical correlation. This result is independent of the distance function adopted, 
affecting all the discordlike measures (see, e.g., a brief discussion for the specific cases of the Bures 
distance~\cite{Bromley} and the measurement-induced disturbance~\cite{Lang:11}). Moreover, by choosing first to optimize the distance 
function for the classical correlation, the nonuniqueness will be moved to the subsequent optimization of the quantum 
correlation. In order to remove those degeneracies for both classical and quantum correlations, we propose a general 
strategy based on an independent optimization procedure. As will be shown, this will lead to the uniqueness of the distance 
functions, providing a consistent theory of correlations.

\section{Distance functions}

Discordlike measures of quantum correlation are typically devised to quantify the disturbance of quantum
states under local measurements. In this sense, even separable states may exhibit quantumness. Quantum 
correlation is part of the total correlation exhibited by a quantum state, which is also composed 
by the classical correlation. Proposals of bona fide measures for quantum, classical, and total 
correlations are expected to obey a set of fundamental criteria~\cite{Brodutch,Modi}: (i) product states have no 
correlations, (ii) all correlations are invariant under local unitary operations, (iii) all correlations 
are non-negative, (iv) total correlations are nonincreasing under local operations, and (v) classical 
states have no quantum correlations. Moreover, an extra assumption has been recently taken as 
necessary~\cite{Hu:12,Tufarelli:12,Piani,Paula}: (vi) quantum correlations are nonincreasing under local operations over 
unmeasured subsystems. This set of fundamental criteria has been used as a guide to validate correlation measures. 
In the following subsections, we will describe possible frameworks to unify correlation definitions, 
which can be established by either measurement-based or measurement-independent approaches.

\subsection{Measurement-based approach}

\label{MB-approach}

In the general method presented by Brodutch and Modi~\cite{Brodutch,Modi} discordlike
measures of quantum, classical, and total correlations of an $n$-partite system in a state $\rho$ are
respectively defined by 
\begin{equation}\label{q}
Q(\rho)=K\left[\rho,M(\rho)\right],
\end{equation}
\begin{equation}\label{c}
C(\rho)=K\left[M(\rho),M(\pi_{\rho}^{m})\right],
\end{equation}
\begin{equation}\label{t}
T(\rho)=K\left[\rho,\pi_{\rho}^{m}\right],
\end{equation} 
where $\pi_{\rho}^{m}=\text{tr}_{\bar{1}}\rho\otimes...\otimes\text{tr}_{\bar{n}}\rho$ represents the product of the local marginals  of $\rho$: the state you would have if all the correlations between the $n$ parts were erased.
$K\left[\rho,\tau\right]$  denotes a real-valued function 
that vanishes for $\rho=\tau$, and $M(\rho)$ is a classical state emerging from a measurement on $\rho$ chosen according to a pre-selected strategy. For an n-partite system $A_1, \cdots, A_n$, 
the classical states assume the form $\chi=\sum_{i}p_{i}\Pi_{A_1}^{\left(i\right)}\otimes...\otimes\Pi_{A_N}^{\left(i\right)}$, where $p_{i}$ is a joint probability distribution and $\{\Pi_{A_j}^{(i)}\}$ is a 
set of orthogonal projectors for subsystem $A_j$. 

It can be shown that a number of correlation quantifiers are compatible with the fundamental criteria of 
correlations listed above. Explicit examples in the class of entropic measures include the mutual information 
$K_I(\rho,\tau)=|I(\rho)-I(\tau)|$, the conditional entropy $K_D(\rho,\tau)=|S(\rho_b|a)-S(\tau_b|a)|$, and the von 
Neumann entropy $K_S(\rho,\tau)=|S(\rho)-S(\tau)|$. In the class of geometric measures, one has the Schatten 1-norm 
(trace norm) $K_G(\rho,\tau)=\Vert\rho-\tau\Vert_1$, with 
$\left\Vert X\right\Vert_{1}={\textrm{Tr}}\left[\sqrt{X^{\dagger}X}\right]$, among others. Concerning $M(\rho)$, 
it is usually  defined as a positive operator-valued measure (POVM) over one or more of the subsystems such that: 
(a) it minimizes the quantum correlation or (b) it maximizes the classical correlation. 

\subsection{Measurement-independent approach}

A measurement-independent approach has also been introduced for different discordlike measures~\cite{Modi:10,Bellomo:12,Aaronson,Bromley}. 
In terms of the generalized function $K$, we can express the quantum, classical, and total correlations in 
this approach respectively by
\begin{equation}\label{ql}
Q'(\rho)=K\left[\rho,\chi_{\rho}\right],
\end{equation}
\begin{equation}\label{cl}
C'(\rho)=K\left[\chi_{\rho},\pi_{\chi_{\rho}}\right],
\end{equation}
\begin{equation}\label{tl}
T'(\rho)=K\left[\rho,\pi_{\rho}\right].
\end{equation}
For the evaluation of the function $K$, $\chi_{\rho}$ denotes the classical state closest to $\rho$ among a 
pre-selected set of classical states $\{\chi\}$, whereas $\pi_{\chi_{\rho}}$ and $\pi_{\rho}$ represent the 
product states closest to $\chi_{\rho}$ and $\rho$, respectively. In comparison with the measurement-based 
formalism, the optimization of Eqs.~(\ref{ql})-(\ref{tl}) may lead to $C'\neq C$ and $T'\neq T$ even in the 
case $Q'=Q$ (which occurs for $\chi_{\rho}=M(\rho)$~\cite{Paula:2,Aaronson}). On the other hand, both approaches 
are equivalent in terms of obeying the fundamental criteria.

\section{Ambiguities}

\begin{figure}[ht!]
\includegraphics[scale=0.32, angle=270]{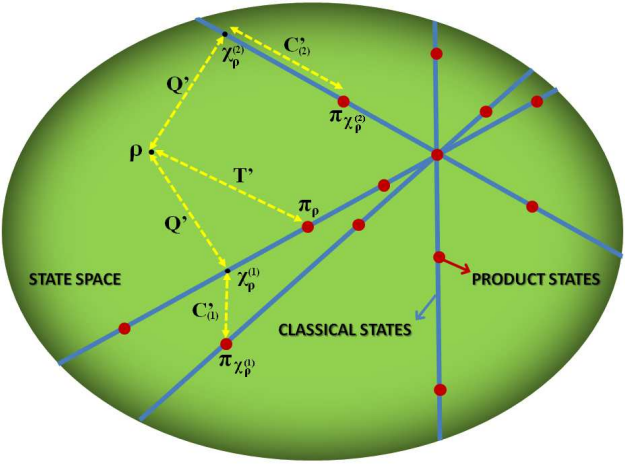} \caption{\label{fig1} (Color online) Schematic picture of degenerate optimized classical states $\chi_{\rho}$ in the 
measurement-independent approach. Note that two distinct classical states $\chi^{(1)}_{\rho}$ and $\chi^{(2)}_{\rho}$ lead to a unique quantum correlation 
$Q^\prime$, but to nonunique classical correlations $C^\prime_{(1)} \ne C^\prime_{(2)}$. }
\end{figure}

Both the measurement-based and the measurement-independent approaches yield well-defined frameworks as long 
as the optimized classical state involved (either $M(\rho)$ or $\chi_{\rho}$) is unique for any given $\rho$. 
However, this assumption is not true in general. Degenerate classical states, which lead to a single value of 
quantum correlation but nonunique values for the classical correlation (and vice-versa in the case of the 
measurement-based framework), are often present in the optimization performed over state space. As an example, 
Fig.~\ref{fig1} provides a sketch of such degeneracies in the measurement-independent formalism.

\begin{figure}[ht!]
\includegraphics[scale=0.42]{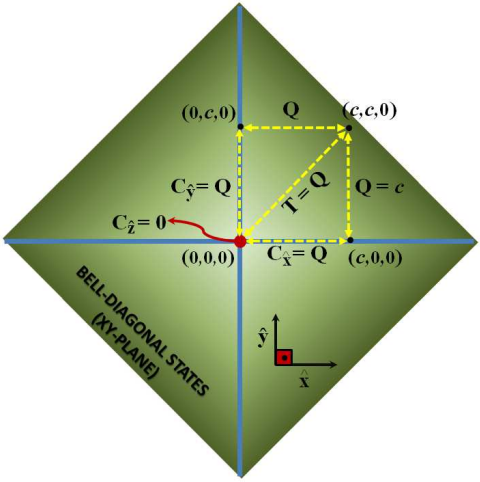} \caption{\label{fig2} (Color online) Schematic picture of a degeneracy in the optimized classical states for a two-qubit Bell-diagonal state. Note that the (blue) solid lines represent the axes of parameter space. 
The quantum state $\rho_{*}$ is characterized by the correlation vector $(c,c,0)$, with equally distant (degenerate) 
classical states $(c,0,0)$, $(0,c,0)$, and $(0,0,0)$.} Note also that the diagonal of the dashed (yellow) square is equal to its sides by measuring distance through 
the trace norm.
\end{figure}

As an explicit illustration of these ambiguities, let us consider a function $K$ based on the Schatten 1-norm 
and a classical state $M(\rho)$ emerging from a projective measurement over one qubit of a two-qubit system so 
that $M(\rho)$ minimizes the quantum correlation $Q$. Then, we have 
\begin{equation}\label{k}
K\left[\rho,\tau\right]=K_{G}\left[\rho,\tau\right]=\left\Vert \rho-\tau\right\Vert_{1}=\sum_{i=1}^{4}\left|\Gamma_{i}\left[\rho-\tau\right]\right|,
\end{equation}
\begin{equation}\label{m}
M(\rho)=M_{\hat{n}}(\rho)=\sum_{j=-,+}\left(\Pi^{j}_{\hat{n}}\otimes \mathbb{I}\right)\rho\left(\Pi^{j}_{\hat{n}}\otimes \mathbb{I}\right),
\end{equation}
where $\{\Gamma_{i}\left[\rho-\tau\right]\}$ represent the eigenvalues of the Hermitian operator $\rho-\tau$ and 
$\Pi^{\pm}_{\hat{n}}=\left(\mathbb{I}\pm \hat{n}\cdot\vec{\sigma}\right)/2$ are projection operators, being 
$\mathbb{I}$ the identity matrix, $\vec{\sigma}=\left(\sigma_{x},\, \sigma_{y},\,\sigma_{z}\right)$ a vector formed 
by Pauli matrices, and $\hat{n}=\left(n_{x},\, n_{y},\,n_{z}\right)$ a unitary vector that minimizes 
Eq.~(\ref{k}).  In this scenario, $Q$ is equivalent to the one-norm geometric quantum discord, which is 
defined by $Q'$ with $\chi_{\rho}$ denoting the classical-quantum state closest to $\rho$~\cite{Paula,Nakano:12}. 
In particular, by taking the two-qubit system as described by a Bell-diagonal state, its density 
operator reads
\begin{equation}\label{rho}
\bar{\rho}=\frac{1}{4}\left[\mathbb{I}\otimes \mathbb{I}+\vec{c}\cdot \left(\vec{\sigma}\otimes\vec{\sigma}\right)\right],
\end{equation}
where $\vec{c}=\left(c_{x},c_{y},c_{z}\right)$ is a three-dimensional vector composed by the correlation functions 
$\{c_{k}\}=\{\langle\sigma_{k}\otimes\sigma_{k}\rangle\}$ ($k=x,\,\,y,$ and $z$). In this case, the optimized classical state, the product of the local marginals of 
$\bar{\rho}$, and the product state closest to the optimized classical state have been computed in Refs.~\cite{Paula:2,Aaronson}, yielding
\begin{equation}\label{mbell}
M_{\hat{n}}(\bar{\rho})=\chi_{\bar{\rho}}=\frac{1}{4}\left(\mathbb{I}\otimes \mathbb{I}+c_{n}\sigma_{n}\otimes\sigma_{n}\right),
\end{equation}
\begin{equation}\label{pirho}
\pi_{\bar{\rho}}^{m}= \frac{1}{4}\left(\mathbb{I}\otimes \mathbb{I}\right),
\end{equation}
\begin{equation}\label{pichirho}
\pi_{\chi_{\bar{\rho}}}= \frac{1}{4}\left(\mathbb{I}\otimes \mathbb{I}+a_{n}\sigma_{n}\otimes\mathbb{I}+b_{n}\mathbb{I}\otimes\sigma_{n}+a_{n}b_{n}\sigma_{n}\otimes\sigma_{n}\right).
\end{equation}
In Eqs.~(\ref{mbell})-(\ref{pichirho}), $\hat{n}$ is such that $|c_n|=\text{max}\{|c_{k}|\}$ (maximum element of the set $\{|c_{k}|\}$), whereas $a_{n}=\mp1\pm\sqrt{1+|c_{n}|}$ and 
$b_{n}=c_{n}a_{n}/|c_{n}|$. Defining $c_+=\text{max}\{|c_{k}|\}$ and $c_{0}=\text{int}\{|c_{k}|\}$ (intermediate element of the set $\{|c_{k}|\}$), Eqs.~(\ref{mbell})-(\ref{pichirho}) 
lead to the following expressions for the quantum and classical correlations~\cite{Paula:2,Aaronson}: 
\begin{equation}\label{qbell}
Q(\bar{\rho})=Q'(\bar{\rho})=c_{0},
\end{equation}
\begin{equation}\label{cbell}
C(\bar{\rho})=c_{+},
\end{equation}
\begin{equation}\label{clbell}
C'(\bar{\rho})=2\left(\sqrt{1+c_{+}} -1\right), 
\end{equation}
where $Q$, $C$, $Q'$, and $C'$ are obtained from Eqs.~(\ref{q}), (\ref{c}), (\ref{ql}), and (\ref{cl}), 
respectively, with the distance function $K$ given in terms of the Schatten 1-norm. 
Now, let us prove that the optimized classical state in Eq.~(\ref{mbell}) is nonunique, implying there are 
multiple values for the classical correlations. In order to provide a concrete example of ambiguity, let us consider a 
restricted class of Bell-diagonal states $\rho_{*}$, which is given by 
\begin{equation}\label{rho2}
\rho_{*}=\frac{1}{4}\left[\mathbb{I}\otimes \mathbb{I}+\vec{c}_{*}\cdot \left(\vec{\sigma}\otimes\vec{\sigma}\right)\right],
\end{equation}
where the correlation vector is $\vec{c}_{*}=\left(c,\,c,\,0\right)$, with $c>0$. In this case, the possible four eigenvalues of the operator $\rho_{*}-M_{\hat{n}}(\rho_{*})$ 
are given by $\{\Gamma_{i}\left[\rho_{*}-M_{\hat{n}}(\rho_{*})\right]\}=\{-\alpha_{-},\alpha_{-},-\alpha_{+},\alpha_{+}\}$, where $\alpha_{\pm}=c(1\pm|n_{z}|)/4$. So, taking 
into account these eigenvalues in Eq.~(\ref{k}), we obtain that the quantum correlation reads
\begin{equation}\label{qbell2}
Q_{\hat{n}}(\rho_{*})=\sum_{i=1}^{4}\left|\Gamma_{i}\left[\rho_{*}-M_{\hat{n}}(\rho_{*})\right]\right|=2(\alpha_{-}+\alpha_{+})=c,
\end{equation}
with $\hat{n}$ an arbitrary unitary vector. This result is in agreement with Eq.~(\ref{qbell}) applied to the state described in Eq.~(\ref{rho2}). However, its independence of 
$\hat{n}$ reveals the existence of an infinity number of optimized classical states associated with a unique value of $Q_{\hat{n}}(\rho_{*})$. Such degeneracy implies into a 
continuum of inequivalent values for the classical correlation. Indeed, the eigenvalues of $M_{\hat{n}}(\rho_{*})-M_{\hat{n}}\left(\pi_{\rho_{*}}^{m}\right)$ are given by 
$\{\Gamma_{i}\left[M_{\hat{n}}(\rho_{*})-M_{\hat{n}}\left(\pi_{\rho_{*}}^{m}\right)\right]\}=\{-\beta,-\beta,\beta,\beta\}$, where $\beta=c\sqrt{1-n_{z}^ 2}/4$. This leads to
\begin{equation}\label{cbell2}
C_{\hat{n}}(\rho_{*})=\sum_{i=1}^{4}\left|\Gamma_{i}\left[M_{\hat{n}}(\rho_{*})-M_{\hat{n}}\left(\pi_{\rho_{*}}^{m}\right)\right]\right|=c\sqrt{1-n_{z}^ 2}
\end{equation}
with $-1\leq n_{z}\leq 1$. Note that Eq.~(\ref{cbell2}) agrees with Eq.~(\ref{cbell}) applied to the state $\rho_{*}$ only in the case $n_{z}=0$, i.e., when $\hat{n}$ lies on the $xy$-plane. 
This includes the particular solutions $\hat{n}=\pm\hat{x}$ or $\pm\hat{y}$ associated with the classical state in Eq.~(\ref{mbell}). The maximal discrepancy occurs for 
$n_{z}=\pm 1$ ($\hat{n}=\pm \hat{z}$), where the classical correlation in Eq.~(\ref{cbell2}) goes to zero. A schematic picture of this situation is shown in Fig.~\ref{fig2}. Concerning the 
measurement-independent framework, we find
\begin{equation}\label{qlbell2}
Q'_{\hat{n}}(\rho_{*})=Q_{\hat{n}}(\rho_{*})=c,
\end{equation}
\begin{equation}\label{clbell2}
C'_{\hat{n}}(\rho_{*})\leq C_{\hat{n}}(\rho_{*}),
\end{equation}
where we have used the relations 
$M_{\hat{n}}(\rho_{*})=\chi_{\rho_{*}}$, $M_{\hat{n}}(\pi_{\rho_{*}}^{m})=\pi_{\rho_{*}}^{m}=\left(\mathbb{I}\otimes \mathbb{I}\right)/4$, and 
$\left\Vert\chi_{\rho_{*}}-\pi_{\chi_{\rho_{*}}}\right\Vert_{1}\leq \left\Vert \chi_{\rho_{*}}-\pi_{\rho_{*}}^{m}\right\Vert_{1}$ into the definitions of quantum and classical correlations. 
From Eqs.~(\ref{mbell}) and (\ref{clbell}), we can derive that 
\begin{equation}
C'_{\pm \hat{x}}(\rho_{*})=C'_{\pm \hat{y}}(\rho_{*})\neq 0, 
\end{equation}
whereas from  Eqs.~(\ref{clbell2}) and (\ref{cbell2}) we obtain 
\begin{equation}
C'_{\pm \hat{z}}(\rho_{*})= 0. 
\end{equation}
This variation of $C'_{\hat{n}}(\rho_{*})$ for distinct choices of $\hat{n}$ is sufficient to show that the nonuniqueness also affects the 
measurement-independent approach.

\section{New framework}

As long as the fundamental criteria of reasonable correlation measures are satisfied, it is plausible to add 
auxiliary strategies to the previous frameworks in order to overcome ambiguities originated from the degeneracies. 
As an example, we could adopt the following modified strategy (see Subsection {\it Measurement-based approach} 
for the original strategy): (a') maximize the classical correlation or (b') minimize the quantum correlation over the degenerate 
subspace. However, depending on the distance function $K$ and for general quantum states, it may not be a trivial 
task to find out the degenerate subspace as well as to optimize over the degenerate classical states. Here, we 
propose a new framework based on the measurement-based approach, which has the advantage of avoiding extra optimization 
over the degenerate subspace. In this new framework, the quantum, classical, and total correlations are independently 
obtained from 
\begin{equation}\label{qll}
Q"(\rho)=K\left[\rho,M_{-}(\rho)\right],
\end{equation}
\begin{equation}\label{cll}
C"(\rho)=K\left[M_{+}(\rho),M_{+}(\pi_{\rho}^{m})\right],
\end{equation}
\begin{equation}\label{tll}
T"(\rho)=K\left[\rho,\pi_{\rho}^{m}\right],
\end{equation} 
where $M_{-}(\rho)$ and $M_{+}(\rho)$ are classical states that minimizes $Q"$ and maximizes $C"$ within a pre-established set of measurements (e.g., orthogonal 
projective measurements), respectively. In this approach, degeneracies in $M_{-}(\rho)$ or $M_{+}(\rho)$ are irrelevant, being sufficient to find out a unique solution for each 
optimal measurement. More specifically, $M_{-}(\rho)$ and $M_{+}(\rho)$ are independently optimized, with no direct connection between them. 
As an application, let us assume $K=K_{G}$, $\rho=\bar{\rho}$, and $M_{\pm}(\rho)=M_{\hat{n}_{\pm}}(\rho)$, i.e., orthogonal projective measurements, 
where $\hat{n}_{-}=\left(n_{x-},n_{y-},n_{z-}\right)$ and $\hat{n}_{+}=\left(n_{x+},n_{y+},n_{z+}\right)$ denote the unitary vectors that minimize $Q"$ and maximize $C"$, respectively. 
From previous analysis, it follows that $\hat{n}_{-}=\hat{n}$ or $M_{\hat{n}_{-}}(\bar{\rho})=M_{\hat{n}}(\bar{\rho})$, with $c_{n}=\text{max}\{c_{k}\}$, such that
\begin{equation}\label{qllbell}
Q"(\bar{\rho})=Q(\bar{\rho})=c_{0}.
\end{equation} 
Concerning the evaluation of the classical correlation, we find $\{\Gamma_{i}\left[M_{\hat{n}_{+}}(\bar{\rho})-M_{\hat{n}_{+}}(\pi_{\bar{\rho}}^{m})\right]\}=\{-\gamma,-\gamma,\gamma,\gamma\}$, 
where we have defined $\gamma=\sqrt{\vec{v}\cdot\vec{u}}/4$ with $\vec{v}=\left(c_{x}^2,c_{y}^2,c_{z}^2\right)$ and $\vec{u}=\left(n_{x+}^2,n_{y+}^2,n_{z+}^2\right)$. These eigenvalues lead to 
\begin{equation}\label{cllbell}
C"(\bar{\rho})=\sum_{i=1}^{4}\left|\Gamma_{i}\left[M_{\hat{n}_{+}}(\bar{\rho})-M_{\hat{n}_{+}}(\pi_{\bar{\rho}}^{m})\right]\right|=\sqrt{\vec{v}\cdot\vec{u}},
\end{equation}
where $\vec{u}$ maximizes Eq.~(\ref{cllbell}) under the conditions $u_{x}+u_{y}+u_{z}=1$ and $0\leq u_{k}\leq 1$. By defining $v_{+}=\text{max}\{v_{k}\}$, $v_{0}=\text{int}\{v_{k}\}$, $v_{-}=\text{min}\{v_{k}\}$, $u_{+}=\text{max}\{u_{k}\}$, $u_{0}=\text{int}\{u_{k}\}$, and $u_{-}=\text{min}\{u_{k}\}$, where max, int, and min, denote maximum, intermediate, and minimum, respectively, we can write
\begin{equation}\label{cond1}
\vec{v}\cdot\vec{u}=v_{x}u_{x}+v_{y}u_{y}+v_{z}u_{z}=v_{+}u_{+}+v_{0}u_{0}+v_{-}u_{-},
\end{equation}
with
\begin{equation}\label{cond2}
u_{x}+u_{y}+u_{z}=u_{+}+u_{0}+u_{-}=1.
\end{equation}
Then, isolating $u_{+}$ in terms of $u_{0}$ and $u_{-}$ in Eq.~(\ref{cond2}) and inserting the resulting expression in Eq.~(\ref{cond1}), we obtain
\begin{equation}
\vec{v}\cdot\vec{u}=v_{+}-\left(v_{+}-v_{0}\right)u_{0}-\left(v_{+}-v_{-}\right)u_{-}\leq v_{+},
\end{equation}
where we have used the relations $v_{+}\geq v_{0}$, $v_{+}\geq v_{-}$, $0\leq u_{0}\leq 1$, and $0\leq u_{-}\leq 1$. Evidently, the maximum value $v_{+}$ of the function $\vec{v}\cdot\vec{u}$ can be 
achieved for $u_{-}=0$, $u_{0}=0$, and $u_{+}=1$, i.e., for $\hat{n}_{+}=\hat{n}_{-}=\hat{n}$. Consequently, $M_{\hat{n}_{+}}(\bar{\rho})=M_{\hat{n}_{-}}(\bar{\rho})=M_{\hat{n}}(\bar{\rho})$. Thus, it follows that \begin{equation}
C"(\bar{\rho})=C(\bar{\rho})=c_{+}.
\label{Cpp}
\end{equation}
Remarkably, the framework introduced here and the measurement-based approach with $M(\rho)$ given by Eq.~(\ref{mbell}) (see Ref.~\cite{Paula:2}) lead to the same expressions for the correlations 
in the particular case of the Bell-diagonal states, even though inequivalent results may appear for more general states. Furthermore, it is also important to emphasize 
that the alternative strategy of further optimization over the degenerate subspace (instead of independent optimization of Q and C) also provides the same results. Indeed, since we have shown 
that there is at least one optimized classical state in common for $M_{-}(\rho)$ and $M_{+}(\rho)$, given by Eq.~(\ref{mbell}), then Eqs.  (\ref{qllbell}) and (\ref{Cpp}) can also be obtained from the 
measurement-based formalism by assuming strategy (a) in combination with (a') or (b) followed by (b').

\section{Conclusions}

In summary, we have identified ambiguities in the definition of either classical or quantum correlations, which 
potentially affect all the approaches used to define discordlike measures. Moreover, we have proposed a new 
framework to avoid such ambiguities by independent optimization of the correlation functions. These results are 
relevant  for a consistent correlation theory and for practical applications of correlation measures, such as in 
quantum many-body systems~\cite{Sarandy:13,Campbell:13}, in the emergence of the pointer basis in open quantum 
systems~\cite{Cornelio:12,Paula:13}, etc. As future challenges, it remains the application of the proposed 
framework for states more general than the Bell-diagonal qubit-qubit states, the investigation of its robustness 
against decoherence, and possible relevance in quantum information tasks. 

\begin{acknowledgments}

We acknowledge Gerardo Adesso for useful discussions. This work is supported by the Brazilian agencies 
CNPq, CAPES, FAPERJ, and the Brazilian National Institute for Science and Technology of Quantum Information (INCT-IQ).

\end{acknowledgments}


\end{document}